\documentclass[12pt]{article}

\usepackage{graphicx}
\begin{document}

\begin{center}
{\bf Remarks on nonsingular models of Hayward and magnetized black hole with rational nonlinear electrodynamics} \\
\vspace{5mm} Sergey Il'ich Kruglov
\footnote{E-mail: serguei.krouglov@utoronto.ca}
\underline{}
\vspace{3mm}

\textit{Department of Physics, University of Toronto, \\60 St. Georges St.,
Toronto, ON M5S 1A7, Canada\\
Department of Chemical and Physical Sciences, University of Toronto Mississauga,\\
3359 Mississauga Road North, Mississauga, Ontario L5L 1C6, Canada} \\
\vspace{5mm}
\end{center}
\begin{abstract}
A Hayward black hole and magnetically charged black hole based on rational nonlinear electrodynamics with the Lagrangian ${\cal L} = -{\cal F}/(1+2\beta{\cal F})$ (${\cal F}$ is a field invariant) are considered. It was shown that the metric function in both models possesses a de Sitter core without singularities as $r\rightarrow 0$. The behavior of the Hawking temperature and the heat capacity in these models are similar. The phase transitions take place when the Hawking temperature has a maximum, and black holes are thermodynamically stable at some event horizon radii when the heat capacity is positive. We show that the source of gravity in the Hayward model is questionable.
\end{abstract}

\vspace{3mm}
\vspace{3mm}

\section{Introduction}

Singularities inside Schwarzschild and Reisner $-$Nordstr\"{o}m black holes (BHs) are problems in General Relativity (GR). But singularities can be considered as nonphysical and due to classical GR.
It is naturally to think that singularities should be avoided.
One of successful regular models of BHs, avoiding singularities, was proposed by Hayward \cite{Hayward}. The Hayward metric function contains a length parameter $l$ to smooth singularities.
Solutions in the theory of GR coupled to nonlinear electrodynamics (NED) have been proposed in
\cite{Pellicer}-\cite{Kruglov}. In these models the physical source of BHs is a NED.
Here, we study regular BH solutions in the Hayward model and magnetically charged BH model based on rational nonlinear electrodynamics (RNED) proposed in \cite{Kr}. The BH thermodynamics and phase transitions in these models are considered.

The paper is organized as follows. In Sect. 2 we study the Hayward BH. Thermodynamics and phase transitions of Hayward BH are investigated in Subsect. 2.1. The BH solutions within RNED are studied in Sect. 3.
We investigate the unitarity and causality principles. The dyonic solution is obtained in Subsect. 3.1.
The regular magnetic BH within RNED is studied in Subsect. 3.2. Thermodynamics and phase transitions of BH are investigated in Subsect. 3.3. The Hawking temperature and the heat capacity are studied. We show that phase transitions take place. Section 4 is a conclusion.

\section{The Hayward BH solution}

We consider the spherically symmetric line element squared which is given by
\begin{equation}
ds^2=-f(r)dt^2+\frac{1}{f(r)}dr^2+r^2(d\vartheta^2+\sin^2\vartheta d\phi^2),
 \label{1}
\end{equation}
where the metric function, within GR, reads
\begin{equation}
f(r)=1-\frac{2m(r)G}{r}.
\label{2}
\end{equation}
The Hayward metric function representing the static regular BH is given by \cite{Hayward}
\begin{equation}
f(r)=1-\frac{2GMr^2}{r^3+2GMl^2},
\label{3}
\end{equation}
where $G$ is the Newton constant, $M$ is the mass parameter, and $l$ is the fundamental length.
Making series expansion of the metric function (3) as $r\rightarrow 0$ and $r\rightarrow\infty$ we obtain
\begin{equation}
f(r)=1-\frac{r^2}{l^2}+\frac{r^5}{2GMl^4}+{\cal O}(r^8)~~~~r\rightarrow 0,
\label{4}
\end{equation}
\begin{equation}
f(r)=1-\frac{2GM}{r}+\frac{4G^2M^2l^2}{r^4}+{\cal O}(r^{-7})~~~~r\rightarrow \infty.
\label{5}
\end{equation}
Equation (4) shows that the metric function in the Hayward model possesses a de Sitter core without singularities. According to Eq. (5) the first two terms reproduce the Schwarzschild behavior of the metric function. But there is not a term containing $r^{-2}$ corresponding to the Reissner$-$Nordst\"{o}m solution of charged BH.
Comparing Eqs. (2) and (3) we obtain the mass function in the Hayward model
\begin{equation}
m(r)=\frac{Mr^3}{r^3+2GMl^2}.
 \label{6}
\end{equation}
We consider the mass function (6) within GR coupled to a NED. In this case the mass function is given by \cite{Bronnikov}
\begin{equation}
m(r)=\int^r_0\rho(r)r^2dr,
\label{7}
\end{equation}
where $\rho$ is the energy density. From Eqs. (6) and (7) we find
\begin{equation}
\rho(r)=\frac{1}{r^2}\frac{dm(r)}{dr}=\frac{6M^2Gl^2}{(r^3+2GMl^2)^2}.
 \label{8}
\end{equation}
In the following we consider only magnetic BH because the electric field (for the models which have Maxwell's weak-field limit) leads to singularities \cite{Bronnikov}.
Then the Lagrangian is given by
\begin{equation}
{\cal L}=-\rho=-\frac{6M^2Gl^2}{(r^3+2GMl^2)^2}.
 \label{9}
\end{equation}
In the case of magnetized BH the field invariant is
\begin{equation}
{\cal F}=\frac{B^2}{2}=\frac{q_m^2}{2r^4},
 \label{10}
\end{equation}
where $B$ is the magnetic field of a monopole and $q_m$ is its magnetic charge. From Eq. (10) we obtain the radius
$r=\sqrt[4]{q_m^2/(2{\cal F})}$ and replacing it in Eq. (9) one finds the electromagnetic Lagrangian corresponding to the Hayward model
\begin{equation}
{\cal L}=-\frac{6M^2Gl^2(2{\cal F})^{3/2}}{(q_m^{3/2}+2^{7/4}GMl^2{\cal F}^{3/4})^2}.
 \label{11}
\end{equation}
One can verify, making use of Eq. (8), that the total magnetic mass of the BH is \cite{Bronnikov}
\begin{equation}
m_M=\int^\infty_0\rho(r)r^2dr=M.
 \label{12}
\end{equation}
Thus, the mass parameter $M$ in the Hayward model is indeed the total magnetic mass when the source of gravity is the electromagnetic field with the Lagrangian (11).
Introducing the dimensional parameter
\begin{equation}
\alpha=2^{7/3}\left(\frac{GMl^2}{q_m^{3/2}}\right)^{4/3}
 \label{13}
\end{equation}
so that $\alpha{\cal F}$ is dimensionless, we represent the Lagrangian (11) in the more convenient form
\begin{equation}
{\cal L}=-\frac{3(\alpha{\cal F})^{3/2}}{2Gl^2(1+(\alpha{\cal F})^{3/4})^2}.
 \label{14}
\end{equation}
It is worth noting that Larangians with the same dependance on the invariant ${\cal F}$ were discuss in \cite{Fan} (see also comments in \cite{Bronnikov3} and \cite{Stuchlik}) and \cite{Nam}.
At the weak-field limit the NED Lagrangian (14) does not become the Maxwell Lagrangian. In addition, the Lagrangian (14) (as well as (11)) contains the gravitation constant $G$ and the dimension parameter $l$ which are unusual for NED models. Thus, the source of gravity in the Hayward model in the framework of GR is questionable.

\subsection{The Hayward BH thermodynamics}

Let us study the black holes thermodynamics and the thermal stability of the Hayward BH. The Hawking temperature is given by
\begin{equation}
T_H=\frac{f'(r_+)}{4\pi},
\label{15}
\end{equation}
where $r_+$ is the event horizon radius which is defined by the bigger root of the equation $f(r_+)=0$. Making use of Eqs. (3) and (15) we obtain the Hawking temperature
\begin{equation}
T_H=\frac{GMr_+(r_+^3-4MGl^2)}{2\pi (r_+^3+2GMl^2)^2}.
\label{16}
\end{equation}
It is convenient to introduce the constant $b=2GMl^2$ and the dimensionless variable $y=r/\sqrt[3]{2b}$ ($y_+=r_+/\sqrt[3]{2b}$). Then Eq. (16) becomes
\begin{equation}
T_H=\frac{\sqrt[3]{b}y_+(y_+^3-1)}{\sqrt[3]{4}\pi l^2(2y_+^3+1)^2}.
\label{17}
\end{equation}
The plot of the dimensionless function $T_H(y_+)l^2/\sqrt[3]{b}$ is depicted in Fig. 1.
\begin{figure}[h]
\includegraphics[height=3.0in,width=3.0in]{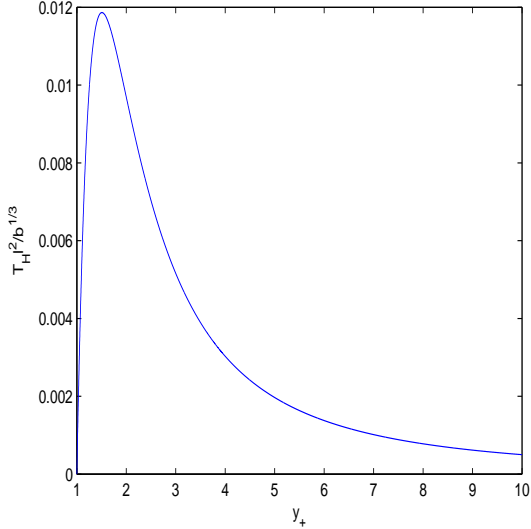}
\caption{\label{fig.1}The plot of the function $T_H(y_+)l^2/\sqrt[3]{b}$.}
\end{figure}
Figure 1 (and Eq. (17)) shows that the Hawking temperature is positive for $y_+>1$. The BH does not exist  \cite{Page} when the temperature is negative, $0<y_+<1$. The maximum of the Hawking temperature takes place when
$\partial T_H/\partial y_+=0$. One finds from Eq. (17)
\begin{equation}
\frac{\partial T_H}{\partial y_+}=\frac{\sqrt[3]{b}(14y_+^3-4y_+^6-1)}{\sqrt[3]{4}\pi l^2(2y_+^3+1)^3}.
 \label{18}
\end{equation}
Making use of Eq. (18) and the condition $\partial T_H/\partial y_+=0$, we obtain that the maximum of the Hawking temperature occurs at $y_+=\sqrt[3]{(7+3\sqrt{5})/4}\approx 1.5$ (see also Fig. 1).
The similar form of the temperature curve for a BH occurs in the models \cite{Myung1}, \cite{Myung}, \cite{Tharanath}.
To study the stability of BH we calculate the heat capacity.
With the help of the Hawking entropy of the BH $S=\mbox{Area}/(4G)=\pi r_+^2/G=\pi y_+^2(2b)^{2/3}/G$ we find the heat capacity \cite{Novikov}
\begin{equation}
C_q=T_H\left(\frac{\partial S}{\partial T_H}\right)_q=\frac{T_H\partial S/\partial y_+}{\partial T_H/\partial y_+}.
\label{19}
\end{equation}
Making use of Eqs. (18) and (19) one obtains
\begin{equation}
C_q=\frac{\sqrt[3]{2^5b^2}\pi y_+^2(y_+^3-1)(2y_+^3+1)}{G(14y_+^3-4y_+^6-1)}.
\label{20}
\end{equation}
According to Eq. (19) the heat capacity possesses a singularity when the Hawking temperature has an extremum ($\partial T_H/\partial y_+=0$). The plot of the heat capacity versus the variable $y_+$ is depicted in Fig. 2.
\begin{figure}[h]
\includegraphics[height=3.0in,width=3.0in]{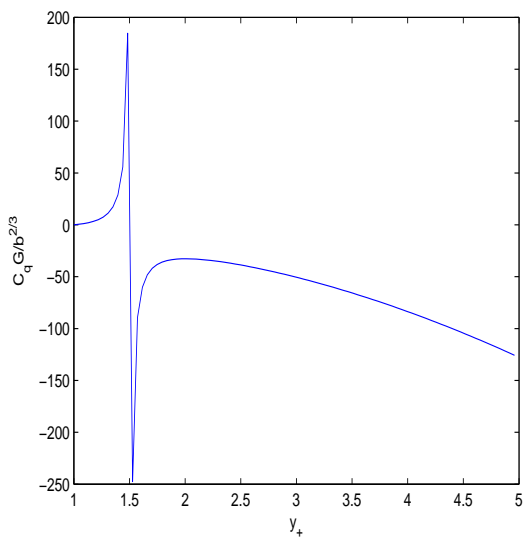}
\caption{\label{fig.2}The plot of the function $GC_q/b^{2/3}$ vs $y_+$. }
\end{figure}
Figure 2 shows that the BH is stable at $1.5>y_+>1$ because the heat capacity is positive.
The singularity in the heat capacity is at the point $y_+\approx 1.5$ where the second-order phase transition occurs.

\section{The BH solution based on RNED}

Let us consider rational NED, proposed in \cite{Kr}, with the Lagrangian
\begin{equation}
{\cal L} = -\frac{{\cal F}}{2\beta{\cal F}+1},
 \label{21}
\end{equation}
where the parameter $\beta$ ($\beta\geq 0$) possesses the dimension of (length)$^4$, ${\cal F}=(1/4)F_{\mu\nu}F^{\mu\nu}=(B^2-E^2)/2$, $F_{\mu\nu}=\partial_\mu A_\nu-\partial_\nu A_\mu$ is the field tensor.
The symmetrical energy-momentum tensor is given by
\begin{equation}
T_{\mu\nu}=-\frac{F_\mu^{~\alpha}F_{\nu\alpha}}{(1+2\beta{\cal F})^{2}}
-g_{\mu\nu}{\cal L}.
\label{22}
\end{equation}
Making use of Eq. (22), we obtain the energy density
\begin{equation}
\rho=T_0^{~0}=\frac{{\cal F}}{1+2\beta{\cal F}}
+\frac{E^2}{(1+2\beta{\cal F})^2}.
\label{23}
\end{equation}
In the valuable theory the general principles of causality and unitarity have to hold. According to the causality principle the group velocity of excitations over the background should be less than the light speed, and then  tachyons are absent in the theory. The absence of ghosts is guaranteed by the unitarity principle. Both principles are satisfied if the following inequalities hold \cite{Shabad2}:
\[
 {\cal L}_{\cal F}\leq 0,~~~~{\cal L}_{{\cal F}{\cal F}}\geq 0,
\] \begin{equation}
{\cal L}_{\cal F}+2{\cal F} {\cal L}_{{\cal F}{\cal F}}\leq 0,
\label{24}
\end{equation}
where ${\cal L}_{\cal F}\equiv\partial{\cal L}/\partial{\cal F}$.
With the help of Eq. (21) we obtain
\[
{\cal L}_{\cal F}= -\frac{1}{(1+2\beta{\cal F})^2},
\]
\begin{equation}
{\cal L}_{\cal F}+2{\cal F} {\cal L}_{{\cal F}{\cal F}}=-\frac{6\beta{\cal F}}{(1+2\beta{\cal F})^3},~~~~
{\cal L}_{{\cal F}{\cal F}}=\frac{4\beta}{(1+2\beta{\cal F})^3}.
\label{25}
\end{equation}
Making use of Eqs. (24) and (25), the principles of causality and unitarity take place if $\beta{\cal F}\geq 0$. For the case $\textbf{E}=0$ Eqs. (24) and (25) are satisfied for any values of the magnetic field. When $\textbf{E}\neq0$, $\textbf{B}\neq0$ (the dyonic case), the restriction $|\textbf{B}|\geq |\textbf{E}|$ is needed.

\subsection{The dyonic solution}

The action of NED coupled with GR is
\begin{equation}
I=\int d^4x\sqrt{-g}\left(\frac{1}{16\pi G}R+ {\cal L}\right),
\label{26}
\end{equation}
where $16\pi G\equiv M_{Pl}^{-2}$, and $M_{Pl}$ is the reduced Planck mass. The Einstein equation is given by
\begin{equation}
R_{\mu\nu}-\frac{1}{2}g_{\mu\nu}R=-8\pi GT_{\mu\nu}.
\label{27}
\end{equation}
By varying action (26) on electromagnetic potentials  one finds the fields equation for electromagnet fields
\begin{equation}
\partial_\mu\left(\sqrt{-g}F^{\mu\nu}{\cal L}_{\cal F}\right)=0.
\label{28}
\end{equation}
Consider the static and spherically symmetric metric with the line element
\begin{equation}
ds^2=-f(r)dt^2+\frac{1}{f(r)}dr^2+r^2(d\vartheta^2+\sin^2\vartheta d\phi^2).
\label{29}
\end{equation}
The general solutions of field equations, found in \cite{Bronnikov1}, \cite{Bronnikov2}, are given by
\begin{equation}
B^2=\frac{q^2_m}{r^4},~~~~E^2=\frac{q_e^2}{{\cal L}^2_{\cal F}r^4},
\label{30}
\end{equation}
where $q_m$ and $q_e$ are the magnetic and electric charges, respectively.
With the help of Eq. (30) we obtain
\begin{equation}
E^2=\frac{q_e^2(1+2\beta{\cal F})^4}{r^4},
\label{31}
\end{equation}
\begin{equation}
\beta{\cal F}=a-b(1+2\beta{\cal F})^4,~~~a=\frac{\beta q^2_m}{2r^4},~~~
b=\frac{\beta q_e^2}{2r^4}.
\label{32}
\end{equation}
We have introduced the dimensionless variables $a$ and $b$. Defining the dimensionless value $x\equiv\beta{\cal F}$, we obtain from Eq. (32) the equation as follows:
\begin{equation}
b(2x+1)^4+x-a=0.
\label{33}
\end{equation}
Making use of the dimensionless variable $t=r/\sqrt[4]{\beta q_m^2}$ and the constant $n=q_m^2/q_e^2$, one finds from Eq. (33) the equation for $z=2x+1$:
\begin{equation}
z^4+t^4z-n-t^4=0.
\label{34}
\end{equation}
The real dyonic solution to Eq. (34) is given by
\[
z=\sqrt{\frac{\sqrt[4]{3}t^4}{4\sqrt[4]{n+t^4}\sqrt{\sinh(\varphi/3)}}-\frac{\sqrt{n+t^4}\sinh(\varphi/3)}
{\sqrt{3}}}-\frac{\sqrt{\sinh(\varphi/3)}\sqrt[4]{n+t^4}}{\sqrt[4]{3}},
\]
\begin{equation}
\sinh(\varphi)=\frac{3^{3/2}t^8}{16(n+t^4)^{3/2}}.
\label{35}
\end{equation}
Putting $n=0$ in Eq. (35) we come to the solution corresponding to the electrically charged BH \cite{Kruglov1}. We find the self-dual solution at $q_e=q_m$ ($a=b$) from Eq. (31). Then $x=0$ (${\cal F}=0$, $E=B$), $E=q/r^2$ ($q\equiv q_e=q_m$) and with the help of Eqs. (7) and (23) we obtain the mass function
\begin{equation}
m(r)=m_0-\int^\infty_r\rho(r)r^2dr=m_0-\frac{q^2}{r},
\label{36}
\end{equation}
where $m_0$ is the mass of the dyonic BH.
Making use of Eq. (2) one finds the metric function
\begin{equation}
f(r)=1-\frac{2m_0G}{r}+\frac{2q^2G}{r^2}.
\label{37}
\end{equation}
The metric function (37) corresponds to the RN solution with $2q^2=q^2_e+q^2_m$. Similar solution was obtained in \cite{Kruglov4} in the logarithmic NED coupled with GR.

\subsection{The magnetic BH within RNED}

Now, we consider the static magnetic BH \footnote{In the paper M.-S. Ma, Ann. Phys. \textbf{362}, 529 (2015) the author also considered the static magnetic BH (but without proper citation) based on NED proposed in \cite{Kr}. However, here we use unitless variables that are more convenient for the analyses of the BH thermodynamics.}. Taking into account that $q_e=0$, ${\cal F}=q_m^2/(2r^4)$, we obtain from Eq. (23) the magnetic energy density
\begin{equation}
\rho_M=\frac{B^2}{2(\beta B^2+1)}=\frac{q_m^2}{2(r^4+\beta q_m^2)}.
\label{38}
\end{equation}
With the help of Eqs. (7) and (38) one finds the mass function
\[
m(x)=\frac{q_m^{3/2}}{8\sqrt{2}\beta^{1/4}}\biggl(\ln\frac{x^2-\sqrt{2}x+1}{x^2+\sqrt{2}x+1}
\]
\begin{equation}
+2\arctan(\sqrt{2}x+1)-2\arctan(1-\sqrt{2}x)\biggr),
\label{39}
\end{equation}
where we introduce the dimensionless variable $x=r/\sqrt[4]{\beta q_m^2}$.
The BH magnetic mass is given by
\begin{equation}
m_M=\int_0^\infty\rho_M(r)r^2dr=\frac{\pi q_m^{3/2}}{4\sqrt{2}\beta^{1/4}}\approx 0.56\frac{q_m^{3/2}}{\beta^{1/4}}.
\label{40}
\end{equation}
Making use of Eq. (2) we obtain the metric function
\[
f(x)=1-\frac{q_mG}{4\sqrt{2\beta}x}\biggl(\ln\frac{x^2-\sqrt{2}x+1}{x^2+\sqrt{2}x+1}
\]
\begin{equation}
+2\arctan(\sqrt{2}x+1)-2\arctan(1-\sqrt{2}x)\biggr).
\label{41}
\end{equation}
As $r\rightarrow \infty$ the metric function (41) becomes
\begin{equation}
f(r)=1-\frac{2m_MG}{r}+\frac{q_m^2G}{r^2}+{\cal O}(r^{-5})~~~~r\rightarrow \infty.
\label{42}
\end{equation}
The correction to the RN solution, according to Eq. (42), is in the order of ${\cal O}(r^{-5})$.
As $r\rightarrow 0$, from Eq. (41) one finds the asymptotic with a de Sitter core
\begin{equation}
f(r)=1-\frac{Gr^2}{\beta}+\frac{Gr^6}{7\beta^2q_m^2}-\frac{Gr^{10}}{11\beta^3q_m^4}+{\cal O}(r^{12})~~~~r\rightarrow 0.
\label{43}
\end{equation}
The solution (41) is regular because as $r\rightarrow 0$ we have $f(r)\rightarrow 1$.
Let us introduce the dimensionless constant $B=q_mG/\sqrt{\beta}$.
Then the horizon radii, that are the roots of the equation $f(r_{+/-})=0$ ($x_{+/-}=r_{+/-}/(\sqrt{q_m}\beta^{1/4})$),  are given in Tables 1.
\begin{table}[ht]
\caption{The horizon radii}
\centering
\begin{tabular}{c c c c c c c c c c  c}\\[1ex]
\hline
$B$ & 3.173 & 3.2 & 3.5 & 4 & 4.5 & 5 & 6 & 7 & 8 \\[0.5ex]
\hline
 $x_-$ & 1.68 & 1.52 & 1.21 & 1.03 & 0.92 & 0.85 & 0.75 & 0.68 & 0.63 \\[0.5ex]
\hline
 $x_+$ & 1.68 & 1.87 & 2.49 & 3.19 & 3.82 & 4.42 & 5.59 & 6.74 & 7.87 \\[0.5ex]
\hline
\end{tabular}
\end{table}
The plot of the metric function (41) is depicted in Fig. 3.
\begin{figure}[h]
\includegraphics[height=3.0in,width=3.0in]{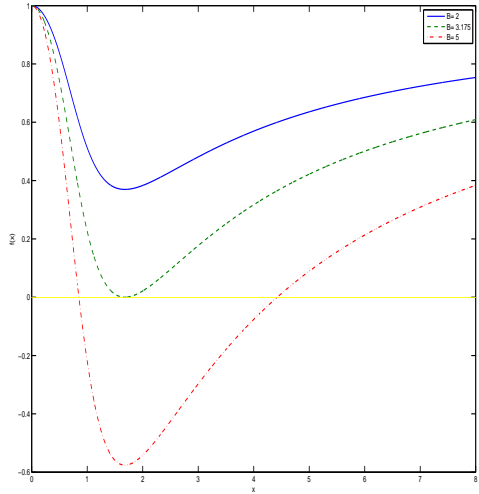}
\caption{\label{fig.3} The plot of the function $f(x)$. The solid curve is for $B=2$, the dashed curve corresponds to $B=3.175$, and the dashed-doted curve corresponds to $B=5$.}
\end{figure}
According to Fig. 3 there are no horizons at $B<3.17$, an extreme horizon occurs at $B\approx 3.173$, and two horizons hold at $B>3.173$.

Making use of Eqs. (22) and (27) at $E=0$, we obtain the Ricci scalar
\begin{equation}
R=8\pi G T_\mu^{~\mu}=\frac{16\pi G\beta q_m^4}{(r^4+\beta q_m^2)^2}.
\label{44}
\end{equation}
The Ricci scalar approaches to zero as $r\rightarrow \infty$ and spacetime becomes flat.

\subsection{The BH thermodynamics with RNED}

Making use of Eqs. (15) and (41) we obtain the Hawking temperature
\begin{equation}
T_H==\frac{1}{4\pi\beta^{1/4}\sqrt{q_m}}\biggl(\frac{1}{x_+}
-\frac{4\sqrt{2}x_+^2}{(1+x_+^4)D}\biggr),
\label{45}
\end{equation}
where
\begin{equation}
D\equiv \ln\frac{x_+^2-\sqrt{2}x_++1}{x_+^2+\sqrt{2}x_++1}
-2\arctan(1-\sqrt{2}x_+)+2\arctan(1+\sqrt{2}x_+).
\label{46}
\end{equation}
The plot of the functions $T_H(x_+)\sqrt{q_m}\beta^{1/4}$ is depicted in Fig. 4.
\begin{figure}[h]
\includegraphics[height=3.0in,width=3.0in]{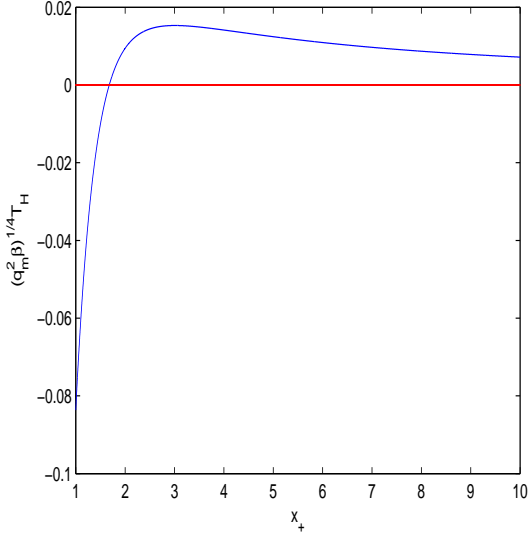}
\caption{\label{fig.4}The plot of the function $T_H\sqrt{q_m}\beta^{1/4}$ vs $x_+$}
\end{figure}
According to Fig. 4 the Hawking temperature is positive for $x_+>1.679 $ and is zero at $x_+\approx 1.679$ .
Making use of the  Hawking entropy of the BH $S=\pi x_+^2q_m\sqrt{\beta}/G$ we find the heat capacity
\begin{equation}
C_q=T_H\left(\frac{\partial S}{\partial T_H}\right)_q=\frac{T_H\partial S/\partial x_+}{\partial T_H/\partial x_+}=\frac{2\pi q_m\sqrt{\beta}x_+T_H}{G\partial T_H/\partial x_+}.
\label{47}
\end{equation}
The plots of the heat capacity versus the variable $x_+$ are depicted in Figs. 5. and 6.
\begin{figure}[h]
\includegraphics[height=3.0in,width=3.0in]{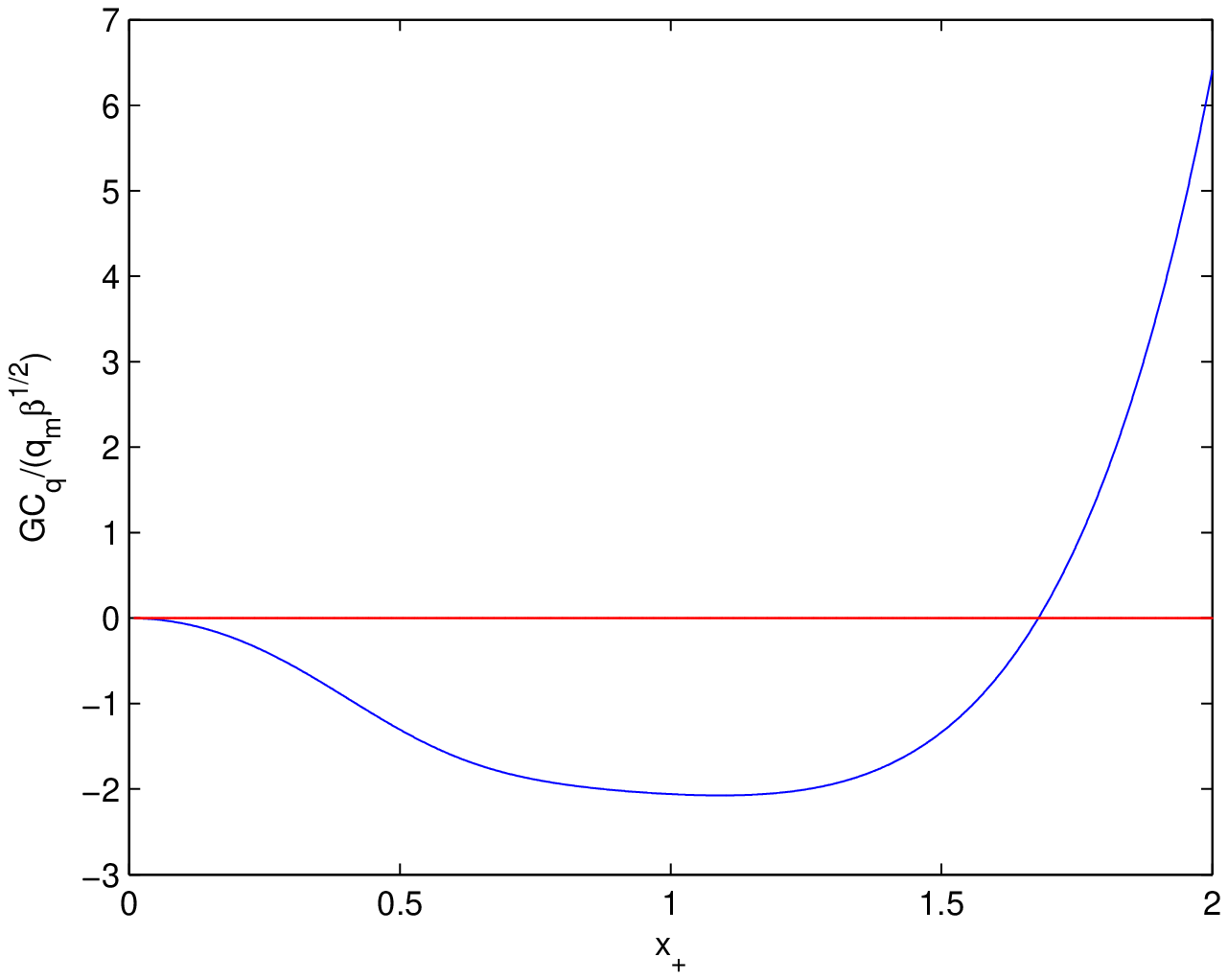}
\caption{\label{fig.5}The plot of the function $GC_q/(q_m^2\beta)^{1/2}$ vs $x_+$.}
\end{figure}
\begin{figure}[h]
\includegraphics[height=3.0in,width=3.0in]{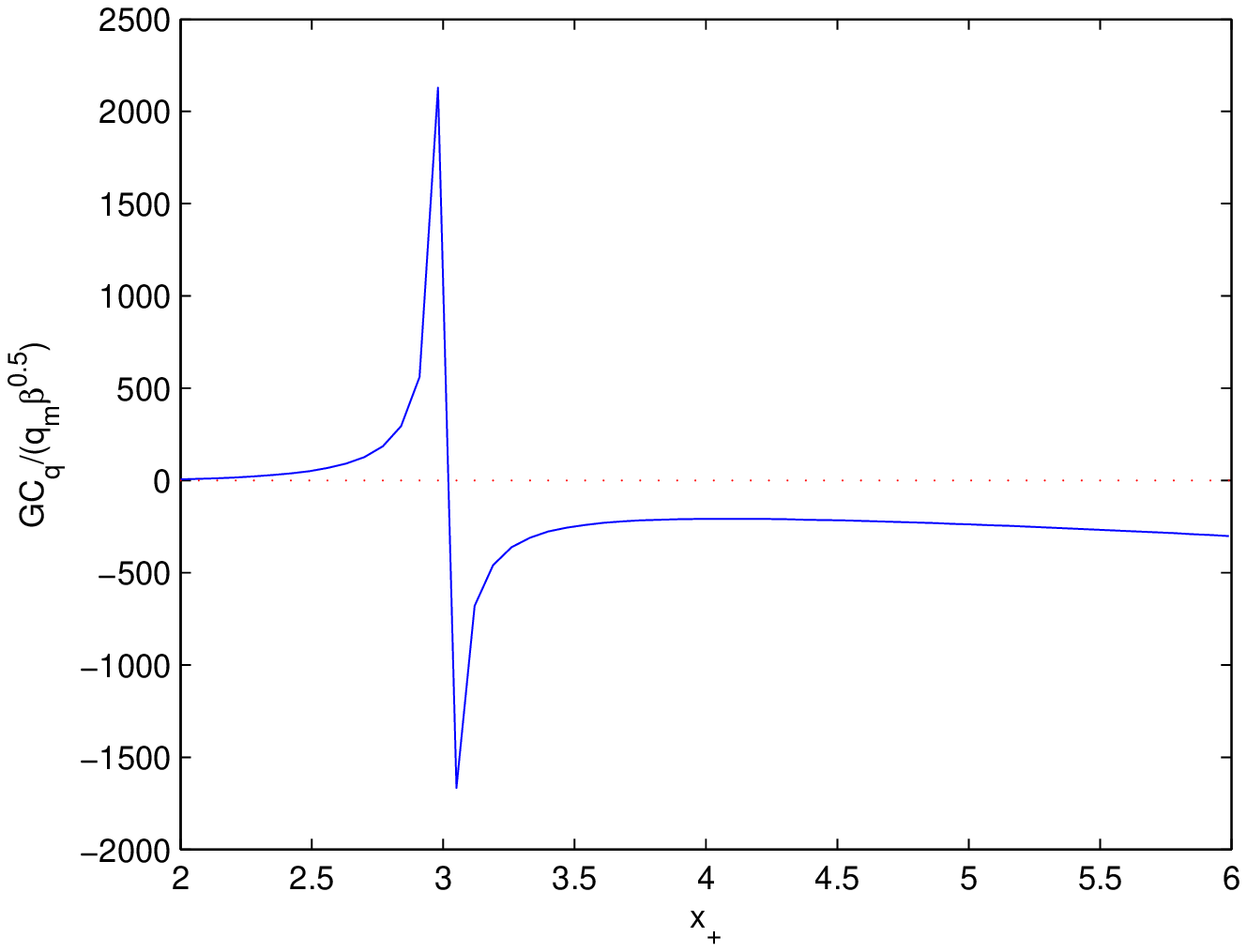}
\caption{\label{fig.6}The plot of the function $GC_q/(q_m^2\beta)^{1/2}$ vs $x_+$.}
\end{figure}
 In accordance with Fig. 5 the BH is unstable at $1.679>x>0$ because the heat capacity is negative.
Figure 6 shows a singularity in the heat capacity at the point $x\approx 3$ where the second-order phase transition occurs. The heat capacity is positive at the range $3>x>1.679$ and the BH is stable.

\section{Conclusion}

The correspondence principle of the Hayward model does not hold as for weak fields NED is not converted into Maxwell's electrodynamics unlike RNED. It was demonstrated that at $|\textbf{B}|\geq |\textbf{E}|$ the principles of causality and unitarity take place in RNED. In RHED the singularity of the electric field at the center of charges is absent and the maximum electric field in the origin is $E(0)=1/\sqrt{\beta}$.
The dyonic and magnetic BHs in GR were studied within RNED. It was shown that in the self-dual case ($q_e=q_m$) the corrections to Coulomb's law and RN solutions are absent. The Ricci scalar does not have the singularity as $r\rightarrow \infty$ and spacetime becomes flat. We shown that the metric function in both models possesses a de Sitter core without singularities as $r\rightarrow 0$. But the Reissner$-$Nordst\"{o}m solution is not reproduced in the Hayward model unlike RNED. In addition, the corresponding NED contains the gravitational constant $G$. Therefore, it is questionable to treat the Hayward model as a solution in GR where the source of gravity is NED.

The thermodynamics and the thermal stability of magnetized BHs were investigated in the Hayward model and RNED coupled to GR. The Hawking temperature, the heat capacity of BHs were calculated and they are similar in both model. It was demonstrated that the heat capacity diverges at some event horizon radii and the phase transitions of the second-order occur. The free parameters in the Hayward model are the mass of the BH $M$ and the fundamental length $l$ wearies in the RNED model the free parameters are the BH magnetic charge $q_m$ and the parameter $\beta$. The source of gravity in the Hayward model is questionable but in our model based on RNED the source of gravity is healthy NED.


\begin{thebibliography}{99}

\bibitem{Hayward} S. A. Hayward, Phys. Rev. Lett. \textbf{96}, 031103 (2006).
\bibitem{Pellicer} R. Pellicer and R. J. Torrence, J. Math. Phys. \textbf{10}, 1718 (1969).
\bibitem{Oliveira} H. P. de Oliveira, Class. Quant. Grav. \textbf{11}, 1469 (1994).
\bibitem{Ayon1} E. Ay\'{o}n-Beato and A. Gar\'{c}ia, Phys. Rev. Lett.  \textbf{80}, 5056 (1998).
\bibitem{Bronnikov0} K. A. Bronnikov, V. N. Melnikov, G. N. Shikin, and K. P. Staniukovich, Ann. Phys.  \textbf{118}, 84 (1979).
\bibitem{Bronnikov} K. A. Bronnikov, Phys. Rev. D \textbf{63}, 044005 (2001).
\bibitem{Bronnikov1}  K. A. Bronnikov, Phys. Rev. Lett. \textbf{85}, 4641 (2000).
 \bibitem{Bronnikov2}   K. A. Bronnikov, G. N. Shikin, and E. N. Sibileva, Grav. Cosmol. \textbf{9}, 169 (2003).
\bibitem{Burinskii} A. Burinskii and S. R. Hildebrandt, Phys. Rev. D \textbf{65}, 104017 (2002).
\bibitem{Diaz} J. Diaz-Alonso and D. Rubiera-Garcia, Phys. Rev. D \textbf{81}, 064021 (2010).
\bibitem{Breton} N. Breton, Gen. Rel. Grav. \textbf{ 37}, 643 (2005).
\bibitem{Novello} M. Novello, S. E. Perez Bergliaffa, and J. M. Salim, Class. Quant. Grav. \textbf{17}, 3821 (2000).
\bibitem{Quiros} R. Garcia-Salcedo, T. Gonzalez, and I. Quiros, Phys. Rev. D \textbf{89}, 084047 (2014).
     	\bibitem{Lemos} J. P. S. Lemos and V. T. Zanchin, Phys. Rev. D \textbf{83}, 124005 (2011).
\bibitem{Balart} L. Balart and E. C. Vagenas, Phys. Rev. D \textbf{90}, 124045 (2014).
\bibitem{Kruglov1}S. I. Kruglov, Phys. Rev. D \textbf{94}, 044026 (2016).
\bibitem{Kruglov} S. I. Kruglov, Annalen Phys. (Berlin) \textbf{528}, 588 (2016); ibid \textbf{529}, 1700073 (2017); Int. J. Mod. Phys. D \textbf{26}, 1750075 (2017); Ann. Phys. \textbf{378}, 59 (2017); ibid \textbf{383}, 550 (2017); Int. J. Mod. Phys. A \textbf{32}, 1700073 (2017); ibid \textbf{33}, 1850023 (2018);  Eur. Phys. J. C \textbf{80}, 250 (2020).
\bibitem{Kr} S. I. Kruglov, Ann. Phys. \textbf{353}, 299 (2015).
\bibitem{Fan} Zhong-Ying Fan and Xiaobao Wang, Phys. Rev. D \textbf{94}, 24027 (2016).
\bibitem{Bronnikov3}K. A. Bronnikov, Phys. Rev. D \textbf{101}, 128501 (2020).
\bibitem{Stuchlik}B. Toshmatov, Z. Stuchlík, B. Ahmedov, Phys. Rev. D \textbf{98}, 028501 (2018)
\bibitem{Nam} C. H. Nam, Eur. Phys. J. C \textbf{78}, 418 (2018).
\bibitem{Page} S. W. Hawking and D. N. Page, Commun. Math. Phys. \textbf{87}, 577 (1983).
\bibitem{Myung1} Y. S. Myung, Y.-W. Kim, and Y.-J. Park, Phys. Lett. B \textbf{656}, 221 (2007).
\bibitem{Myung} Y. S. Myung, Y.-W. Kim, and Y.-J. Park, Gen. Rel. Grav. \textbf{41}, 1051 (2009).
\bibitem{Tharanath} R. Tharanath, J. Suresh, and V. C. Kuriakose, Gen. Rel. Grav. \textbf{47}, 46 (2015).
\bibitem{Novikov} I. D. Novikov and V. P. Frolov, Physics of Black Holes (Kluver Academic Publishers, 1989).
\bibitem{Shabad2}A. E. Shabad and V. V. Usov, Phys. Rev. D \textbf{83}, 105006 (2011).5).
\bibitem{Bronnikov1} K. A. Bronnikov, Grav. Cosmol. \textbf{23}, 343 (2017).
\bibitem{Bronnikov2} K. A. Bronnikov, Int. J. Mod. Phys. D \textbf{27}, 1841005 (2018).
\bibitem{Kruglov4} S. I. Kruglov, Grav. Cosmol. \textbf{25}, 190 (2019).

\end{thebibliography}
\end{document}